\documentclass{article}
\usepackage[utf8]{inputenc}
\usepackage[margin=1 in]{geometry}

\usepackage{lineno}
\usepackage{placeins}
\usepackage{bbold}
\usepackage{amssymb, amsmath, amsthm, mathtools, color}
\usepackage{booktabs}       
\usepackage{multirow}
\usepackage{multicol}        
\usepackage{longtable}
\usepackage{array}
\usepackage{rotating}
\usepackage{float}
\usepackage{subcaption} 
\usepackage{epsfig}
\usepackage{url}
\usepackage{graphicx}
\usepackage{csquotes}
\usepackage{enumerate}
\usepackage{amsmath}
\usepackage{listings}
\usepackage{color}
\usepackage{physics}
\usepackage{listings}
\usepackage{comment}
\usepackage[title]{appendix}
\usepackage{natbib}
\usepackage{hyperref}
\usepackage{soul}
\usepackage{pdfpages}
\usepackage{authblk}

\title{\LARGE\bfseries Using Importance Sampling to Estimate $p$-values in\\
       All-Subset Meta-Analysis, with Applications to\\
       Single-Cell eQTL Mapping}

\author[1]{Samuel Anyaso-Samuel}
\author[1]{Thong Luong}
\author[1]{Fei Qin}
\author[1]{Jiyeon Choi}
\author[1]{Kai Yu}
\author[1,$\ast$]{Paul S. Albert}
\author[1,$\ast$]{Jianxin Shi}

\affil[1]{\small Division of Cancer Epidemiology and Genetics, National Cancer Institute, Rockville, MD, U.S.A.}
\date{}
\begin{document}

\maketitle

\begin{abstract}
Pooling genome-wide association studies of multiple related traits can substantially increase power for detecting genetic variants with pleiotropic effects. ASSET, which exhaustively searches all subsets of studies for association signals, has been widely used to detect modest effects and improve interpretability. Under a normality assumption, ASSET computes p-values via an analytic approximation that accounts for multiple testing. However, this approximation has been evaluated only in limited scenarios and for p-values no smaller than $10^{-3}$. A systematic assessment in the extreme tail is therefore needed, yet naïve Monte Carlo methods would require prohibitively many simulations. We develop a computationally efficient importance-sampling (IS) algorithm that provides accurate ASSET p-value estimates for both independent and overlapping studies, achieving substantial efficiency gains over naïve Monte Carlo, particularly for very small p-values. Using IS, we show that ASSET's analytic approximation is highly accurate across nearly the entire p-value range when normality holds. In contrast, when normality is violated (due to small sample sizes, low-frequency variants, or non-normal traits), ASSET p-values can be inflated or deflated by orders of magnitude, whereas our IS approach remains accurate. We illustrate the method through applications to single-cell eQTL mapping using peripheral blood mononuclear cells from the OneK1K cohort and lung cells from a Korean population.
\end{abstract}

\section{Introduction}
Many genetic applications require pooling data, either summary statistics or individual-level data, from studies of related traits or conditions to improve statistical power for detecting pleiotropic associations with modest effect sizes and to enhance biological interpretability. Examples include (i) genome-wide association studies (GWAS) of multiple related traits (e.g., cancers \citep{fehringer2016cross} or autoimmune diseases \citep{cotsapas2011pervasive}) that aim to identify genetic variants associated with more than one disease, (ii) GWAS of a disease in multiple studies of various ancestry that aim to identify risk variants across multiple populations \citep{mahajan2022multi}, and (iii) expression quantitative trait locus (eQTL) analyses in single-cell transcriptomic studies that seek to detect variants associated with gene expression across multiple cell types \citep{yazar2022single, luong2026single, natri2024cell, soskic2022immune, chen2025deciphering}. Fixed-effect meta-analysis is commonly used to pool such data, but it can suffer substantial power loss when a variant is associated with only one or a small subset of traits or conditions.

To address this limitation, ASSET \citep{bhattacharjee2012subset} was developed to efficiently combine evidence across studies by exhaustively searching all nonempty subsets of traits. By allowing for heterogeneous effects across studies, ASSET provides robust power for detecting pleiotropic signals that may be missed by traditional meta-analytic approaches, albeit at the cost of a multiple testing adjustment. ASSET has since been widely used in genetic studies to identify modest associations \citep{li2015meta, kar2016genome, jee2025dissecting, qi2024genome} that would otherwise remain undetected due to limited sample sizes in individual studies.

A critical challenge in applying ASSET is accurately evaluating the significance of its test statistic while accounting for the multiple testing induced by the exhaustive subset search. Under a multivariate normal assumption, the original ASSET authors developed an analytic approximation for the p-value using a discrete local maxima (DLM) method \citep{taylor2007maxima}, and demonstrated through simulations that this approximation is reasonably accurate when p-values $> 0.001$ and sample sizes are sufficiently large to justify asymptotic normality of the individual test statistics. However, because computation of the ASSET statistic involves evaluation over all $2^{M}-1$ nonempty subsets (where $M$ denotes the number of studies or traits), it becomes computationally prohibitive to directly assess the accuracy of the analytic approximation for extremely small p-values (e.g., $10^{-8}$) across a wide range of settings, including varying sample sizes, phenotype distributions, and allele frequencies. Such extreme significance levels are routinely required for declaring genome-wide, study-wide, or phenome-wide significance in genetic applications.

In this manuscript, we develop importance sampling (IS) algorithms to efficiently assess the significance of the ASSET statistic for two main purposes: (i) to verify the accuracy of the ASSET analytic approximation over a wide range of p-values when the normality assumption holds, and (ii) to provide an alternative and computationally efficient method for estimating ASSET p-values when the normality assumption fails, a scenario that is particularly consequential for extreme tail probabilities. IS is well-suited for estimating small tail probabilities compared with naïve Monte Carlo simulation, as it generates samples under an alternative probability measure that makes rare events occur more frequently, thereby dramatically reducing estimation variance. IS has been successfully applied in various contexts, including evaluation of the significance of scan statistics with applications in genetics and medical image analysis \citep{naiman2001computing, malley2003comprehensive}, estimation of genome-wide significance for score statistics in linkage studies with fully informative \citep{angquist2004using} or partially informative markers \citep{shi2007importance}, as well as in GWAS based on case-control designs \citep{kimmel2006fast}.

The remainder of the manuscript is organized as follows. In Section \ref{sec: ASSET_norm}, we briefly review the ASSET methodology and existing IS algorithms for general Gaussian random processes, and then develop our IS algorithms for ASSET under normally distributed individual statistics, allowing for both independent and overlapping studies. In Section \ref{sec: ASSET_nonnorm}, we extend the proposed IS framework to genetic studies with small or moderate sample sizes, where the normality assumption may not hold. We illustrate these methods in the context of single-cell eQTL mapping, with the goal of identifying genetic variants associated with gene expression across some or all cell types. In Section \ref{sec: simulations}, we conduct extensive simulations to assess the accuracy of the proposed IS estimators for ASSET p-values, and systematically evaluate the accuracy of the analytic p-value approximation over a wide range of significance levels. In particular, we performed simulations by conditioning on real gene expression data from the single-cell eQTL study in the OneK1K cohort \citep{yazar2022single} to investigate performance when the normality assumption does not hold. In Section \ref{sec: real_data}, we applied the IS algorithm to a single-cell eQTL study of lung tissues in a Korean population. Finally, in Section \ref{sec: discussion}, we briefly discuss some further extensions.

\section{Importance sampling for ASSET assuming normality}\label{sec: ASSET_norm}
\subsection{ASSET statistic and its significance}\label{sec: ASSET_overview}
Consider $M$ independent studies testing the association between a genetic variant and a continuous trait. For the $m$th study, let $z_{m}$ be a score statistic and $n_m$ be the sample size. Assuming a fixed effect in a subset (denoted as $A$) of $M$ studies, we perform a meta-analysis to pool the data in the subset
\begin{align}
 Z_{A} = \sum_{m \in A}{\sqrt{w_{m}(A)}\,z_{m}},
\end{align}
where the optimal weights (in the absence of covariate adjustment) are $w_{m}(A) = n_{m}/\sum_{l \in A}n_{l}$. These weights remain approximately optimal when covariate adjustment is similar across studies \citep{skol2006joint}. To test the global null hypothesis (that no study has a signal) with maximum power, the ASSET method performs pooled analysis exhaustively in all $2^{M} - 1$ subsets and takes the maximum value:
$$Z_\text{ASSET} = \max_{A}{|Z_{A}|}.$$

Given an observed value $b = Z_\text{ASSET}$, the overall significance is $p = P_{0}(\max_{A}\left| Z_{A} \right| > b)$, where $P_{0}$ indicates that none of the studies has a signal. In the original article \citep{bhattacharjee2012subset}, the authors derived a lower bound
\begin{align}
p \ge p_\text{DLM} = \sum_A \int_b^{\infty}
2 \prod_{j=1}^{J}
P_0\left(|Z_{A \pm j}| < z \mid Z_A = z\right)
\phi(z)\,dz,
\end{align}
based on a discrete local maxima (DLM) method \citep{taylor2007maxima} under a multivariate normal assumption for $(z_1,\ldots,z_M)$. Here, $J$ denotes the number of neighbors of $A$, and $A \pm j$ denotes the $j$th neighbor of the current subset $A$ obtained by adding the $j$th study if it is not already included and dropping it otherwise. They demonstrated through simulation that the DLM approximation adequately controlled the Type~I error rate for moderate significance levels (up to 0.001 in Supplementary Tables 4--5 in \cite{bhattacharjee2012subset}) and sufficiently large sample sizes, where asymptotic normality of the study-specific statistics is justified.

For pooling case-control studies, the optimal weights are calculated as $w_{m}(A) = n_{m}^\text{eff}/\sum_{l \in A}n_{l}^\text{eff}$, where $n_{m}^\text{eff} = n_{m}\phi_{m}(1 - \phi_{m})$ is the effective sample size, $n_{m}$ is the total number of subjects, and $\phi_{m}$ is the fraction of cases in the $m$th study.

\subsection{Importance sampling for estimating $P_{0}(\max_{t}z_{t} > b)$}\label{sec: IS_overview}
We briefly review an IS algorithm for estimating the tail probability of a discrete Gaussian random process. For a discrete Gaussian random process $\{ z_{1},\cdots,z_{M}\}$ with $z_{t}\sim N(0,1)$ and $r_{st} = \text{cor}(z_{s},z_{t})$, we are interested in estimating $p = P_{0}\left( \max_{t}z_{t} > b \right)$ for a large threshold value of $b$. Here, the notation $P_{0}$ indicates that the Gaussian process has marginal distribution $N(0,1)$ for all random variables. Under naïve Monte Carlo with $K$ simulations, the standard error is $\sqrt{p(1-p)/K} \approx \sqrt{p/K}$ for small $p$. To achieve a relative error of $10\%$, one requires approximately $100/p$ simulations.

IS can be used for efficiently estimating $p$. To proceed, one defines a new probability measure by a standard embedding of $P_{0}$ into an exponential family \citep{siegmund1976importance}
\begin{align}
dP_{\xi,\tau}\left( z_{1},\cdots,z_{M} \right) = \exp\left( \xi z_{\tau} - \xi^{2}/2 \right)dP_{0}.
\end{align}
Under this alternative probability measure,
$E_{\xi,\tau}(z_{\tau}) = \xi$ and
$E_{\xi,\tau}(z_{t}) = \xi r_{\tau t}$ for $t \neq \tau$, i.e., the non-zero signal in the random process is driven by the random variable indexed by $\tau$. Based on the likelihood ratio identity \citep{pollak1998new, siegmund2000tail}, we have
\begin{align}
    P_{0}\left( \max_{t}z_{t} > b \right) = \frac{1}{M}\sum_{\tau}{E_{\xi,\tau}\left( \frac{M}{\sum_{t = 1}^{M}{\exp(\xi z_{t} - \xi^{2}/2)}}\,\mathcal{I}\!\left\{\max_{{t}} z_{{t}} > b\right\}
\right)}.
\end{align}
An IS algorithm can be developed to estimate $p$ \citep{angquist2004using, shi2007importance} based on (4):

\begin{enumerate}
\def\labelenumi{(\arabic{enumi})}
\item Choose $\xi$;
\item For $k=1,\ldots,K$, randomly select $\tau \in \{ 1,\cdots,M\}$, and simulate
  $\left( z_{1},\cdots,z_{M} \right)$ under the alternative probability measure $P_{\xi,\tau}$, i.e., $\text{MVN}((\mu_{1},\cdots,\mu_{M}),\Sigma)$ with
  $\mu_{t} = r_{t\tau}\xi$.
  \item Calculate
$$
\delta_k = \frac{M}{\sum_{t=1}^{M}\exp\{\xi z_t - \xi^{2}/2\}} \mathcal{I}\!\left\{\max_t z_t > b\right\}.
$$
\item $p$ is estimated as
  ${\widehat{p}}_{\xi} = \frac{1}{K}\sum_{k = 1}^{K}\delta_{k}$ with
  $\mathrm{Var}(\widehat{p}_{\xi}) = \widehat{\mathrm{Var}}(\delta_k)/K.$
\end{enumerate}
The computational efficiency of the IS algorithm is defined as $\mathcal{E}=\frac{p(1-p)/K}{\text{Var}\left( \delta_{k} \right)/K} = \frac{p}{\text{Var}\left( \delta_{k} \right)}$, which represents the ratio of the number of naïve Monte Carlo simulations to the number of IS simulations required to achieve the same estimation accuracy. Choosing the optimal $\xi$ to precisely minimize $\mathrm{Var}(\widehat{p}_{\xi})$ depends on the correlation matrix $\Sigma$ and is analytically challenging. A practical and near-optimal choice is $\xi = b$, which shifts $E_{\xi,\tau}(z_{\tau}) = b$ and makes $P_{\xi,\tau}(\max_t z_t > b)$ approximately $1/2$ (see also analytic justification in \cite{shi2007importance}).

\subsection{Importance sampling for ASSET in independent studies}\label{sec: ASSET_norm_ind}
Following the IS algorithm in Section~\ref{sec: IS_overview}, we consider two approaches to estimating $p(b) = P_{0}(\max_{A} |Z_{A}| > b)$ and develop the second into a practical IS algorithm.

The first approach explicitly calculates $\mathrm{cor}(Z_{A_{1}},Z_{A_{2}})$ for any two subsets $A_{1}$ and $A_{2}$ \citep{bhattacharjee2012subset} and directly applies the Gaussian-process IS algorithm in Section~\ref{sec: IS_overview} to estimate $p(b)$. Although conceptually straightforward, this requires simulating the process $\{Z_A\}$ of dimension $2^{M}-1$, which is computationally intensive. Moreover, this formulation depends critically on the multivariate normal assumption and does not extend naturally to non-Gaussian settings.

The second approach simulates $(z_{1},\cdots,z_{M})$ under an alternative probability measure. To proceed, we define a probability measure
$$
dP_{\xi,A}\left( z_{1},\cdots,z_{M} \right) = \exp\!\left( \xi Z_{A} - \frac{\xi^{2}}{2} \right)dP_{0}\left( z_{1},\cdots,z_{M} \right).$$
Under the new probability measure, only studies in subset $A$ carry a nonzero signal of equal effect size, while studies outside $A$ remain null. A sign $s \in \{+1,-1\}$ is drawn uniformly alongside the driving subset $A_0$. Because the tilting parameter is the signed scalar $s\xi$, the likelihood-ratio identity takes the form
\begin{align}
P_{0}\!\left( \max_{A}|Z_{A}| > b \right)
= \frac{1}{2(2^{M} - 1)}\sum_{A_0}\sum_{s\in\{+1,-1\}}
E_{s\xi,\,A_{0}}\!\left(
\frac{2(2^{M} - 1)}
     {\sum_{A}\bigl[e^{\xi Z_{A} - \xi^{2}/2}
                  + e^{-\xi Z_{A} - \xi^{2}/2}\bigr]}
\,\mathcal{I}\!\left\{\max_A |Z_A| > b\right\}
\right),
\end{align}
with the near-optimal choice $\xi = b$. Under the probability measure for subset $A$, suppose $E_{\xi,A}(z_m) = \beta \sqrt{n_m}$ for $m \in A$,
and $E_{\xi,A}(z_m)=0$ for $m\notin A$. Then
$$E_{\xi,A}(Z_A) = \beta \sqrt{\sum_{m\in A} n_m}.$$
Therefore, choosing $\beta = b/\sqrt{\sum_{m\in A} n_m}$ ensures $E_{\xi,A}(Z_A)=b$.
This yields the following IS algorithm for estimating $p(b)$ for ASSET under independent studies and the normality assumption.

\begin{enumerate}
\def\labelenumi{(\arabic{enumi})}
\item For each simulation $k=1,\ldots,K$, randomly select a subset $A_k$ from the $2^{M}-1$ nonempty subsets and independently draw a sign $s_k \in \{+1, -1\}$ with equal probability.
\item Simulate independent
  $(z_{1},\cdots,z_{M})$ under $P_{s_k\xi,A_k}$, where
\[
z_m \sim
\begin{cases}
N\!\left( s_k\,b\,\dfrac{\sqrt{n_m}}{\sqrt{\sum_{l\in A_k} n_l}},\,1\right), & m\in A_k,\\[4pt]
N(0,1), & m\notin A_k.
\end{cases}
\]
\item Calculate
$$\delta_k = \frac{2(2^{M}-1)}{\sum_{A}\bigl[\exp\{\xi Z_A - \xi^2/2\} \mathbf{{}+{}} \exp\{-\xi Z_A - \xi^2/2\}\bigr]}
\,\mathcal{I}\!\left\{\max_A \mathbf{|}Z_A\mathbf{|} > b\right\}.$$
\item The estimator is
$\widehat{p}_{\xi} = \frac{1}{K}\sum_{k=1}^{K} \delta_k$,
with $\mathrm{Var}(\widehat{p}_{\xi}) = \widehat{\mathrm{Var}}(\delta_k)/K$.
\end{enumerate}
For case-control studies, we replace the sample size $n_{m}$ with the effective sample size $n_{m}^\text{eff}$.

\subsection{Importance sampling for overlapping studies assuming normality}\label{sec: ASSET_norm_cor}
When studies share subjects, the elements of $(z_{1},\ldots,z_{M})$ are correlated. Let $\Sigma$ denote the $M\times M$ correlation matrix of $(z_1,\ldots,z_M)$. For a subset $A$, let $\Sigma_A$ denote the corresponding principal submatrix, $z_A$ the subvector indexed by $A$, and let
$N_A = (\sqrt{n_m})_{m\in A}$ be the vector of square roots of sample sizes. Under the fixed-effect meta-analysis framework, the subset meta-analysis statistic is
$$Z_{A} = \frac{N_{A}^{\top}{\Sigma_A^{-1}}z_{A}}{\sqrt{N_{A}^{\top}{\Sigma_A^{-1}}N_{A}}},$$
which follows $N(0,1)$ asymptotically under the global null hypothesis.

As in the independent case, one may proceed in two ways. The first approach computes $\mathrm{cor}(Z_A,Z_B)$ explicitly \citep{bhattacharjee2012subset} and applies the IS algorithm in Section~\ref{sec: IS_overview} to the $(2^{M}-1)$-dimensional process $\{Z_A\}$. The second approach is to simulate $(z_{1},\cdots,z_{M})$ under an alternative probability measure. Similarly, we define a probability measure $dP_{\xi,A}( z_{1},\cdots,z_{M}) = \exp( \xi Z_{A} - \xi^{2}/2)dP_{0}$ by requiring: (1) only studies in subset $A$ have the signal with the same effect size and (2) choosing $\xi$ such that $E_{\xi,A}(Z_{A}) = b$, or equivalently $\beta = b/\sqrt{N_{A}^{\top}{\Sigma_A^{-1}}N_{A}}$. An IS algorithm analogous to the independent case can then be implemented under this probability measure.

\section{Importance sampling when normality assumption fails}\label{sec: ASSET_nonnorm}
The IS algorithms developed in Section~\ref{sec: ASSET_norm} rely on the test statistics following a normal distribution. This is a reasonable assumption for large studies where the test statistics are asymptotically normal. In genetic studies with small sample sizes, however, the normality assumption may be violated, necessitating modifications to the IS framework to account for non-Gaussian behavior. Such scenarios arise, for example, in (i) eQTL mapping in single-cell RNA sequencing studies aimed at identifying variants associated with gene expression across multiple cell types, (ii) meta-analysis of multiple studies of rare diseases in the same population with small to moderate sample sizes, and (iii) meta-analysis of rare diseases in populations of various ancestry. In this section, we illustrate the proposed IS algorithm in the context of single-cell eQTL mapping (i), with extensions to the case-control settings in (ii) and (iii) deferred to Section~B.3 of the Supplementary Material.

\subsection{Single-cell eQTL assuming expression data are independent across cell types}\label{sec: sc-eQTL_ind}

We consider testing the association between a genetic variant and the expression level of a given gene across $C$ cell types in a study of $N$ subjects. Let $y_{ic}$ denote the expression level for cell type $c$ in subject $i$, and let $Y=\{y_{ic}\}_{1\le i\le N,\,1\le c\le C}$ denote the resulting expression matrix. Throughout, we assume that $y_{ic}$ has been centered and scaled to have mean zero and unit variance across subjects for each cell type. Let $g_i$ denote the genotype for subject $i$, coded as $0, 1, 2$ copies of the minor allele. For simplicity, we assume expression measurements are independent across cell types and no covariate adjustment. The full version of IS with a realistic cross-cell-type correlation structure is presented in Section~\ref{sec: sc-eQTL_cor}, and covariate adjustment is discussed in Section~B.1 of the Supplementary Material.

The score statistic for testing the association for cell type $c$ is
$z_c = \sum_{i=1}^{N} y_{ic}(g_i - 2f)/\sqrt{N}\,\sigma_g,$
where $f$ denotes the allele frequency, assumed known, and $\sigma_g^2 = 2f(1-f)$. 
For a given subset $A \subseteq \{1,\ldots,C\}$, the fixed-effect meta-analysis statistic is $Z_{A} = \sum_{c \in A}z_{c}/\sqrt{|A|}$, because the score statistic for each cell type has approximately the same noncentrality parameter under the same effect size. Using the expression for $z_c$, it can be rewritten as
\begin{equation}
    Z_{A} = \frac{\sum_{c \in A}z_{c}}{\sqrt{|A|}} = \sum_{i = 1}^{N}\frac{\sum_{c \in A}y_{ic}}{\sqrt{N|A|}\,\sigma_{g}}(g_{i} - 2f) = \sum_{i = 1}^{N}\omega_{i,A}\left( g_{i} - 2f \right),
\label{eqn: ASSET_nonnorm}
\end{equation}
as a weighted sum of centered genotypic values, with weights
$$\omega_{i,A} = \frac{\sum_{c \in A}y_{ic}}{\sqrt{N|A|}\,\sigma_{g}}$$
determined entirely by the observed expression data. The ASSET statistic is defined as $Z_\text{ASSET} = \max_{A}|Z_{A}|$ and the significance is $p(b) = P_{0}(\max_{A}|Z_{A}| > b)$ for an observed value $b = Z_{\mathrm{ASSET}}$.

To ensure that p-value calculation is robust to the observed expression values, we compute p-values conditional on the observed expression data, i.e., $p(b\mid Y)=P_0\!\left(\max_A |Z_A|>b \mid Y\right)$. Under this framework, the randomness arises solely from the genotypic values $(g_1,\ldots,g_N)$. If $p(b\mid Y)$ were estimated using a naïve Monte Carlo approach, one could either permute the observed genotypes or simulate genotypes from a binomial distribution with allele frequency $f$, assuming Hardy--Weinberg equilibrium (HWE).

To develop an IS estimator, we define an alternative probability measure that conditions on the observed expression data $Y$. Specifically, we define
\begin{align}
  dP_{\xi,A}\!\left( g_1,\ldots,g_N \mid Y \right)
= \exp\!\left\{ \xi Z_A - \phi_A(\xi) \right\}
\, dP_0\!\left( g_1,\ldots,g_N \right),
\label{eqn: prob_measure_nonnorm}
\end{align}
where
$\phi_A(\xi) = \log E_0\!\left[ \exp\!\left( \xi Z_A \right) \mid Y \right]$
is the cumulant generating function (CGF) of $Z_A$ conditional on $Y$. Under the probability measure $P_{\xi,A}$, the conditional expectation of $Z_A$ satisfies $E_{\xi,A}( Z_A \mid Y ) = \phi_A'(\xi)$. The tilting parameter $\xi$ for each subset $A$ is chosen as the solution to $\phi_A'(\xi) = b$.
Using the linear representation of $Z_A$, we compute $\phi_A(\xi)$ as
\begin{align}
\phi_A(\xi) &= \log E_0\!\left[ \exp\!\left\{ \xi \sum_{i=1}^{N} \omega_{i,A} (g_i - 2f)
\right\} \mid Y\right] \notag \\
&= \sum_{i=1}^{N} \log E_0\!\left[ \exp\!\left\{ \xi \omega_{i,A}
(g_i - 2f) \right\} \right],
\end{align}
where, $E_0\!\left[\exp\{\xi \omega_{i,A}(g_i-2f)\}\right] = (1-f)^2 e^{-2f\xi\omega_{i,A}} + 2f(1-f)e^{(1-2f)\xi\omega_{i,A}} + f^2 e^{(2-2f)\xi\omega_{i,A}}.$

The IS mixture estimator proceeds analogously to Section~\ref{sec: ASSET_norm_ind}: a driving subset $A_k$ and a sign $s_k \in \{+1,-1\}$ are each drawn uniformly at random for each simulation, with $\xi$ solved at $+b$ (for $s_k = +1$) or at $-b$ (for $s_k = -1$). Because the CGF $\phi_A(\xi)$ is not symmetric in $\xi$ for general data-dependent weights $\omega_{i,A}$, two per-subset tilting parameters are required: $\xi_{A,+}$ satisfying $\phi_A'(\xi_{A,+}) = +b$ and $\xi_{A,-}$ satisfying $\phi_A'(\xi_{A,-}) = -b$. Both are obtained by Newton's method. The IS weight denominator then involves $2(2^C-1)$ terms:
$$\delta_k = \frac{2(2^C-1)}{\sum_{A}\bigl[\exp\{\xi_{A,+} Z_A - \phi_A(\xi_{A,+})\} + \exp\{\xi_{A,-} Z_A - \phi_A(\xi_{A,-})\}\bigr]}\,\mathcal{I}\!\left\{\max_A|Z_A|>b\right\}.$$

Under the tilted probability measure $P_{\xi,A}$, the conditional distribution of each genotype $g_i$ satisfies
\begin{align}
  P_{\xi,A}\!\left( g_i = k \mid Y \right)
\propto \exp\!\left\{ \xi \omega_{i,A} (k - 2f) \right\}
P_0\!\left( g_i = k \right),
  \label{eqn: tilt_nonnorm}
\end{align}
where $P_0(g_i = k)$ takes values $(1-f)^2$, $2f(1-f)$, and $f^2$ for $k = 0, 1, 2$, respectively, under HWE. This representation enables efficient simulation of genotypes for all $N$ subjects conditional on $Y$ and allele frequency $f$.

\subsection{Single-cell eQTL with correlation across cell types}\label{sec: sc-eQTL_cor}

In practice, expression measurements across cell types are potentially correlated, and this correlation propagates to the vector of score statistics. We retain the conditional inference framework of Section~\ref{sec: sc-eQTL_ind}: inference is conditional on $Y$, so the only source of randomness under the null is the genotype vector $(g_1,\ldots,g_N)$.

Under $H_0$ and conditional on $Y$, one can show that the conditional covariance matrix of $\mathbf{z} = (z_1,\ldots,z_C)^\top$ equals the sample correlation matrix of the expression vectors:
\begin{equation}
\mathrm{Cov}_0(\mathbf z\mid Y)=\widehat\Sigma_Y, \qquad (\widehat\Sigma_Y)_{cc'}=\frac{1}{N}\sum_{i=1}^N y_{ic}y_{ic'}.
\label{eq:cov_z_givenY}
\end{equation}

For a nonempty subset $A\subseteq\{1,\ldots,C\}$, let $\mathbf{z}_A$ denote the subvector of $\mathbf{z}$ and let $\Sigma_A$ be the corresponding principal submatrix of $\widehat\Sigma_Y$. Since all cell types share the same subjects, the effective sample size is the same for all cell types, and under a fixed-effect alternative with a common effect across all cell types in $A$, the subset meta-analysis statistic is
\begin{equation}
Z_A = \frac{\mathbf 1_A^\top \Sigma_A^{-1}\mathbf z_A}{\sqrt{\mathbf 1_A^\top \Sigma_A^{-1}\mathbf 1_A}},
\label{eq:ZA_GLS}
\end{equation}
with $E_0(Z_A\mid Y)=0$ and $\mathrm{Var}_0(Z_A\mid Y)=1$. As shown in Section~B.2 of the Supplementary Material, $Z_A$ admits the same individual-level linear representation as \eqref{eqn: ASSET_nonnorm},
$Z_A = \sum_{i=1}^N \omega_{i,A}(g_i - 2f)$. Because $Z_A$ retains this linear genotype structure, the IS algorithm of Section~\ref{sec: sc-eQTL_ind} applies directly with only the weights changed. Covariate adjustment extends analogously.

\section{Simulations}\label{sec: simulations}
We conducted extensive simulations to (1) assess whether the proposed IS algorithm provides accurate p-value estimates by comparing it with naïve Monte Carlo simulations, and (2) assess the accuracy of the analytic approximations used in the original ASSET method across the full p-value range by benchmarking against IS results. Simulations were first performed under normally distributed test statistics for both independent and correlated studies. We then conducted simulations in the context of single-cell eQTL mapping using realistic cell-type-specific gene expression data from the OneK1K cohort \citep{yazar2022single} to examine how the accuracy of the ASSET approximation varies with gene expression distribution, minor allele frequency (MAF), and sample size.

\subsection{Importance sampling for ASSET assuming normality}
\begin{figure}[ht]
\centering
\begin{minipage}{0.4\linewidth}
  \centering
  \includegraphics[width=\linewidth]{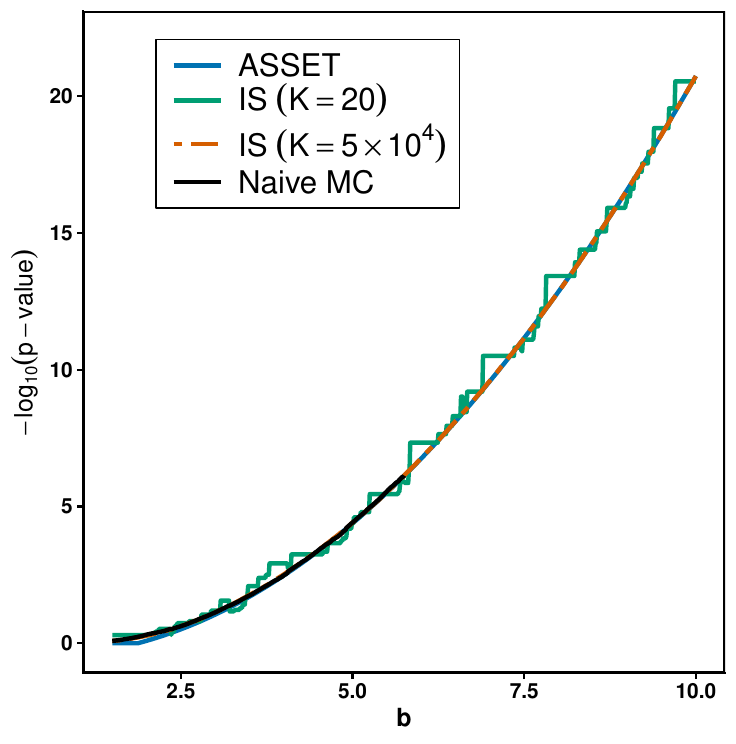}
\end{minipage}
\begin{minipage}{0.4\linewidth}
  \centering
  \includegraphics[width=\linewidth]{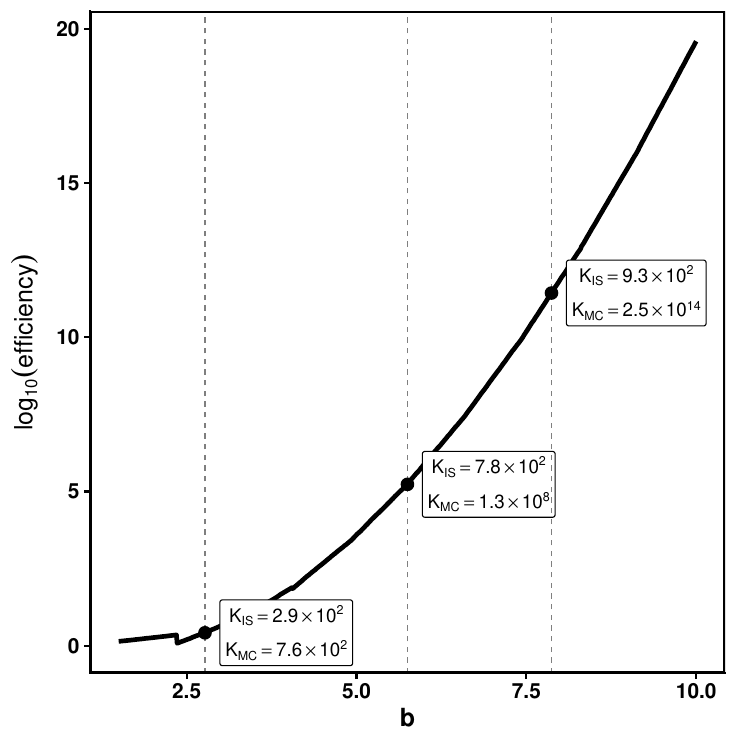}
\end{minipage}
\vspace{-0.3cm}
\caption{Simulation results for with $M=7$ independent studies. 
\textbf{Left:} Curves of $-\log_{10}\{p(b)\}$ versus $b$ obtained using IS, ASSET analytic approximation, and naive Monte Carlo (MC). 
\textbf{Right:} Nominal efficiency $\mathcal{E}=K_{\mathrm{MC}}/K_{\mathrm{IS}}$ of IS over naive MC. Here, $K_{\mathrm{IS}}$ and $K_{\mathrm{MC}}$ are the numbers of simulations required to achieve a standard error of $0.1\times p(b)$. }
\label{fig: independent_studies}
\end{figure}

We first considered the scenario of independent studies under the global null, where the study-level score statistics satisfy $z_m \sim N(0,1)$ independently for $m=1,\ldots,M$. Simulations were conducted for $M=7$ and $M=10$ studies, requiring evaluation of $2^M - 1$ nonempty subsets when computing the ASSET statistic. For each setting, we performed $K = 5 \times 10^4$ IS simulations to obtain accurate estimates of tail probabilities. To achieve a relative standard error of $\mathrm{se}(\widehat{p})/p = 0.1$, naïve MC requires approximately $100/p$ simulations. For computational feasibility, we therefore restricted MC-based estimation to p-values greater than $7.5\times 10^{-6}$, corresponding to a maximum of $1.4 \times 10^7$ MC simulations. Results are summarized in Table~\ref{tab: independent_studies} and Figure~\ref{fig: independent_studies}.

First, the estimated tail probabilities from the IS and naïve MC approaches were in close agreement, confirming that the proposed IS algorithm provides accurate p-value estimates for the ASSET statistic.

Second, the analytic approximation in ASSET based on the DLM method was highly accurate across the full range of $b$ values, supporting its application in large genetic studies. Notably, prior evaluations of the ASSET approximation have focused primarily on moderate p-values (e.g., $p > 10^{-3}$) under asymptotic normality, whereas accuracy at genome-wide significance levels ($5\times 10^{-8}$) and beyond has not been systematically examined. Our results demonstrate that the ASSET approximation remains highly accurate even for extremely small p-values.

Third, we computed the number of simulations required to achieve a relative standard error of $\mathrm{se}(\hat{p})/p = 0.1$ for both IS and naïve MC, denoted by $K_{\mathrm{IS}}$ and $K_{\mathrm{MC}}$, respectively. While $K_{\mathrm{IS}}$ increased only gradually with $b$, $K_{\mathrm{MC}}$ grew rapidly, demonstrating the substantially greater efficiency $\mathcal{E} = K_{\mathrm{MC}}/K_{\mathrm{IS}}$ of IS for estimating very small p-values. For example, when $M=7$ and $b=4.48$, achieving $10\%$ relative precision requires about 550 IS simulations versus $2.2 \times 10^5$ MC simulations, corresponding to an efficiency gain of $\mathcal{E}\approx 4.2 \times 10^2$. This advantage increases sharply with $b$: at $b=7.03$, the estimated efficiency reaches approximately $10^8$. Even modest values of $K_{\mathrm{IS}}$ (e.g., $K_{\mathrm{IS}}=20$) yield reasonably accurate estimates for extremely small p-values (Figure~\ref{fig: independent_studies}).

Finally, note that $P(\max_A |Z_A| > b) = 1 - P(-b \le Z_A \le b\ \text{for all } A)$, and $P(-b \le Z_A \le b\ \text{for all } A)$ can, in principle, be approximated using the R function \texttt{pmvnorm} applied to the $(2^{M}-1)$-dimensional joint distribution of the $\{Z_A\}$ statistics under the multivariate normal assumption, with the corresponding correlation matrix. However, this approach has two major limitations. First, the dimension $(2^{M}-1)$ becomes computationally prohibitive for \texttt{pmvnorm}, even for moderate values such as $M=10$. Second, \texttt{pmvnorm} controls the absolute error for estimating $P(-b \le Z_A \le b\ \text{for all } A)$. When $b$ is large, this quantity is close to one, and even a highly accurate estimate on this scale can yield a numerically unstable estimate of $1 - P(-b \le Z_A \le b\ \text{for all } A)$, leading to substantial loss of precision or collapse to zero due to numerical underflow.

In addition, we performed simulations for the scenario in which $(z_1,\ldots,z_M)$ are correlated, corresponding to meta-analysis studies that may share subjects. Results are presented in Supplementary Table~C.1 and Supplementary Figure~C.1, which show similar accuracy and efficiency gains for the IS approach.

\begin{table}[!ht]
\centering
\caption{Estimation of $p(b)=P_0(\max_A |Z_A|>b)$. The quantities $\widehat{p}_{\mathrm{MVN}}$, $\widehat{p}_{\mathrm{ASSET}}$, $\widehat{p}_{\mathrm{IS}}$, and $\widehat{p}_{\mathrm{MC}}$ denote p-value estimates from the R function \texttt{pmvnorm} applied to the $(2^M-1)$-dimensional joint distribution of the $\{Z_A\}$, the ASSET analytic approximation, importance sampling, and naïve Monte Carlo, respectively; standard errors for IS and MC are shown in parentheses. $K_{\mathrm{IS}}$ and $K_{\mathrm{MC}}$ are the numbers of simulations required to achieve $\mathrm{se}(\hat p)/p = 0.1$, and $\mathcal{E}=K_{\mathrm{MC}}/K_{\mathrm{IS}}$ denotes the nominal efficiency gain. IS estimates use $5\times10^4$ iterations; MC is run using $K_{\mathrm{MC}}$ simulations where feasible, and entries are omitted (``--'') where $p$ is too small for MC to be practical. }
\label{tab: independent_studies}
\centering
\resizebox{1.0\textwidth}{!}{%
\begin{tabular}[t]{ccccccccc}
\toprule
$M$ & $b$ & $\widehat{p}_{\text{MVN}}$ & $\widehat{p}_{\text{ASSET}}$ & $\widehat{p}_{\text{IS}}$ & $\widehat{p}_{\text{MC}}$ & $K_{\text{IS}}$ & $K_{\text{MC}}$ & $\mathcal{E}$\\
\midrule
 & 3.63 & 1.1e-02 & 1.3e-02 & 1.1e-02 (1.1e-04) & 1.2e-02 (3.4e-04) & 4.3e+02 & 8.7e+03 & 2.0e+01\\

 & 4.48 & 3.9e-04 & 4.6e-04 & 4.3e-04 (4.6e-06) & 4.4e-04 (2.1e-05) & 5.5e+02 & 2.2e+05 & 4.2e+02\\

 & 5.33 & 2.8e-06 & 7.7e-06 & 7.5e-06 (8.5e-08) & 7.3e-06 (2.7e-07) & 6.4e+02 & 1.4e+07 & 2.1e+04\\

 & 6.18 & 1.1e-07 & 5.9e-08 & 5.9e-08 (7.1e-10) & - & 7.3e+02 & 1.7e+09 & 2.3e+06\\

 & 7.03 & 3.8e-11 & 2.1e-10 & 2.1e-10 (2.7e-12) & - & 8.4e+02 & 4.8e+11 & 5.7e+08\\

 & 7.88 & 4.3e-14 & 3.7e-13 & 3.6e-13 (5.0e-15) & - & 9.3e+02 & 2.8e+14 & 3.0e+11\\

 & 8.73 & 0.0e+00 & 3.0e-16 & 2.9e-16 (4.2e-18) & - & 1.0e+03 & 3.4e+17 & 3.3e+14\\

\multirow[t]{-8}{*}{\raggedleft\arraybackslash 7} & 9.58 & 0.0e+00 & 1.2e-19 & 1.2e-19 (1.8e-21) & - & 1.1e+03 & 8.4e+20 & 7.6e+17\\
\cmidrule{1-9}
 & 3.63 & - & 4.1e-02 & 3.1e-02 (3.2e-04) & 3.1e-02 (5.4e-04) & 5.2e+02 & 3.3e+03 & 6.2e+00\\

 & 4.48 & - & 1.8e-03 & 1.5e-03 (1.8e-05) & 1.5e-03 (3.8e-05) & 6.8e+02 & 6.8e+04 & 9.6e+01\\

 & 5.33 & - & 3.5e-05 & 3.1e-05 (3.9e-07) & 3.2e-05 (5.6e-07) & 7.6e+02 & 3.1e+06 & 4.2e+03\\

 & 6.18 & - & 3.1e-07 & 2.8e-07 (3.6e-09) & - & 8.5e+02 & 3.6e+08 & 4.2e+05\\

 & 7.03 & - & 1.2e-09 & 1.2e-09 (1.6e-11) & - & 9.2e+02 & 8.6e+10 & 9.3e+07\\

 & 7.88 & - & 2.3e-12 & 2.2e-12 (3.2e-14) & - & 1.0e+03 & 4.5e+13 & 4.5e+10\\

 & 8.73 & - & 2.0e-15 & 2.0e-15 (2.9e-17) & - & 1.1e+03 & 5.1e+16 & 4.7e+13\\

\multirow[t]{-8}{*}{\raggedleft\arraybackslash 10} & 9.58 & - & 8.4e-19 & 8.4e-19 (1.3e-20) & - & 1.1e+03 & 1.2e+20 & 1.0e+17\\
\bottomrule
\end{tabular}
}
\end{table}

\subsection{Simulations conditional on expression data from a single-cell eQTL study of the OneK1K cohort}\label{sec: OneK1K_sim}
In this section, we conducted simulations to evaluate the performance of the IS algorithm for a single-cell eQTL study conditional on observed expression values and to compare it with the analytic approximation used in ASSET.

Simulations were based on real gene expression data from the OneK1K PBMC single-cell eQTL study. We used log-CPM-transformed expression values, standardized to mean zero and unit variance across subjects, for four genes (\texttt{RPL17}, \texttt{RPS26}, \texttt{TPT1}, and \texttt{CDKN1A}) across the $C=7$ most abundant cell types. Simulations were performed for MAF $\in \{0.01, 0.02, 0.05, 0.1, 0.5\}$ and for sample sizes $N=100$, 200, and 981 (the full OneK1K cohort).

The four selected genes exhibit qualitatively distinct empirical expression distributions: approximately normal (\texttt{TPT1}), bimodal (\texttt{RPL17}), strongly bimodal with a sharp dominant mode (\texttt{RPS26}), and zero-inflated (\texttt{CDKN1A}). Histograms and Q--Q plots are shown in Supplementary Figure~C.5. Simulation results are summarized in Table~\ref{tab: oneK1K_studies} and Figure~\ref{fig: oneK1K_TPT1_CDKN1A} for genes \texttt{TPT1} and \texttt{CDKN1A}, and in Supplementary Figure~C.4 for genes \texttt{RPL17} and \texttt{RPS26}. In all figures, IS estimates are obtained using the range approximation, in which a single tilting parameter $\xi$ is reused over a neighborhood of $b$ values.

First, the IS and naïve MC estimates are in close agreement across all scenarios, confirming that IS provides accurate conditional p-value estimates for all settings considered.

Second, sample size, MAF, and gene expression distribution substantially influence p-values at a fixed threshold $b$, often by orders of magnitude. For example, with \texttt{RPL17} expression and $b=5.25$, the p-value decreases from $1.1\times10^{-4}$ at $N=100$ to $3.2\times10^{-5}$ at $N=200$ and $3.9\times10^{-6}$ at $N=981$ (MAF $= 0.01$). At $N=100$, increasing MAF from 0.01 to 0.1 reduces the p-value to $1.5\times10^{-6}$ (and further at higher MAF). P-values also vary markedly across genes: with $N=100$, MAF $= 0.01$, and $b=5.25$, the estimated p-value is $6.3\times10^{-5}$ for \texttt{RPS26}, $6.5\times10^{-5}$ for \texttt{TPT1}, and $1.3\times10^{-4}$ for \texttt{CDKN1A}. These differences become even more pronounced at larger $b$ (Figure~\ref{fig: oneK1K_TPT1_CDKN1A} and Supplementary Figure~C.4), underscoring that accurate p-value computation must account for expression distribution, MAF, and sample size.

Third, $\widehat{p}_{\mathrm{ASSET}}$ can deviate substantially from the IS estimates. These deviations are most pronounced for small sample sizes, and for genes with highly non-normal expression. With moderate sample sizes ($N=100$ or 200) and rare alleles (MAF $\in \{0.01, 0.02, 0.05\}$), the ASSET analytic approximation tends to yield anti-conservative p-values, particularly at large $b$. In contrast, for very common alleles (e.g., MAF $= 0.5$), ASSET tends to be conservative. The discrepancy between ASSET and IS decreases as sample size increases, reflecting improved normal approximations of the test statistics. For example, when $N=981$, the ASSET analytic approximation agrees closely with IS.

Finally, across all settings where MC is reported, IS achieves the targeted relative precision with substantially fewer simulations (Table~\ref{tab: oneK1K_studies}). For example, for \texttt{RPL17} at $N=981$ and MAF $= 0.5$, $p(b)\approx 6.8\times 10^{-7}$ at $b=5.25$, so MC requires $\sim$$3\times 10^{8}$ simulations for a $10\%$ relative standard error, whereas IS attains this precision with $\sim$900 simulations. The efficiency gain is even more pronounced for rarer variants: for \texttt{RPS26} at MAF $= 0.10$ and $b=5.25$, $p(b)$ is on the order of $10^{-8}$--$10^{-9}$, where MC becomes computationally impractical.

Overall, these results show that the proposed IS algorithm, conditional on observed expression data, provides accurate p-value estimates across a range of expression distributions and allele frequencies, and remains computationally viable in tail regimes where naïve MC is prohibitive. At the same time, the comparison highlights that ASSET can be inaccurate when expression distributions are highly non-Gaussian and/or sample sizes are small, whereas its estimates converge to IS estimates in large samples where the normal approximation is better justified.

\begin{table}
\centering
\caption{Estimation of $p(b\mid Y)=P_0(\max_A |Z_A|>b\mid Y)$ conditional on the expression values of four genes in a single-cell eQTL study of the OneK1K cohort. Reported are the p-value estimates from IS ($\widehat{p}_{\mathrm{IS}}$), ASSET analytic approximation ($\widehat{p}_{\mathrm{ASSET}}$), and naïve MC ($\widehat{p}_{\mathrm{MC}}$), together with $\mathcal{E} = K_{\mathrm{MC}}/K_{\mathrm{IS}}$, the nominal efficiency gain of IS over naïve MC. Results are shown for random subsamples ($N=100, 200$) and the full cohort ($N=981$). IS uses 50{,}000 simulations; MC uses $K_{\mathrm{MC}}$ simulations calibrated to achieve $\mathrm{se}(\hat{p})/p = 0.1$.}
\label{tab: oneK1K_studies}
\centering
\resizebox{1.0\textwidth}{!}{%
\begin{tabular}[t]{lcc|cccc|cccc|cccc}
\toprule
\multicolumn{3}{c}{ } & \multicolumn{4}{c}{$N=100$} & \multicolumn{4}{c}{$N=200$} & \multicolumn{4}{c}{$N=981$} \\
\cmidrule(l{3pt}r{3pt}){4-7} \cmidrule(l{3pt}r{3pt}){8-11} \cmidrule(l{3pt}r{3pt}){12-15}
Gene & MAF & $b$ & $\widehat{p}_{\text{ASSET}}$ & $\widehat{p}_{\text{IS}}$ & $\widehat{p}_{\text{MC}}$ & $\mathcal{E}$ & $\widehat{p}_{\text{ASSET}}$ & $\widehat{p}_{\text{IS}}$ & $\widehat{p}_{\text{MC}}$ &  $\mathcal{E}$ & $\widehat{p}_{\text{ASSET}}$ & $\widehat{p}_{\text{IS}}$ & $\widehat{p}_{\text{MC}}$ &  $\mathcal{E}$\\
\midrule
 &  & 4.35 & 6.9e-05 & 3.2e-05 & 3.0e-05 & 3.7e+03 & 6.6e-05 & 3.8e-05 & 3.6e-05 & 3.5e+03 & 5.2e-05 & 5.3e-05 & 5.3e-05 & 2.8e+03\\

 & \multirow[t]{-2}{*}{\raggedleft\arraybackslash 0.50} & 5.25 & 9.9e-07 & 2.9e-07 & 2.7e-07 & 2.8e+05 & 9.5e-07 & 4.0e-07 & 3.9e-07 & 2.7e+05 & 7.2e-07 & 7.1e-07 & 6.8e-07 & 1.5e+05\\

 &  & 4.35 & 6.9e-05 & 6.8e-05 & 6.9e-05 & 1.3e+03 & 6.6e-05 & 5.6e-05 & 5.7e-05 & 2.3e+03 & 5.2e-05 & 5.7e-05 & 5.8e-05 & 2.4e+03\\

 & \multirow[t]{-2}{*}{\raggedleft\arraybackslash 0.10} & 5.25 & 9.9e-07 & 1.4e-06 & 1.5e-06 & 3.4e+04 & 9.5e-07 & 9.4e-07 & 9.5e-07 & 1.0e+05 & 7.2e-07 & 8.6e-07 & 8.6e-07 & 1.2e+05\\

 &  & 4.35 & 6.9e-05 & 9.3e-04 & 9.3e-04 & 1.2e+02 & 6.6e-05 & 4.1e-04 & 4.2e-04 & 2.7e+02 & 5.2e-05 & 1.2e-04 & 1.3e-04 & 9.6e+02\\

\multirow[t]{-6}{*}{\raggedright\arraybackslash RPL17} & \multirow[t]{-2}{*}{\raggedleft\arraybackslash 0.01} & 5.25 & 9.9e-07 & 1.1e-04 & 1.1e-04 & 7.2e+02 & 9.5e-07 & 3.2e-05 & 3.2e-05 & 2.2e+03 & 7.2e-07 & 4.0e-06 & 3.9e-06 & 2.1e+04\\
\cmidrule{1-15}
 &  & 4.35 & 8.8e-05 & 9.5e-06 & 9.0e-06 & 2.0e+04 & 4.3e-05 & 2.2e-05 & 2.3e-05 & 7.7e+03 & 3.2e-05 & 3.3e-05 & 3.0e-05 & 5.5e+03\\

 & \multirow[t]{-2}{*}{\raggedleft\arraybackslash 0.50} & 5.25 & 1.3e-06 & 6.2e-08 & 6.0e-08 & 2.6e+06 & 5.8e-07 & 2.2e-07 & 2.1e-07 & 5.8e+05 & 4.1e-07 & 4.1e-07 & 3.9e-07 & 3.4e+05\\

 &  & 4.35 & 8.8e-05 & 2.6e-05 & 2.4e-05 & 5.2e+03 & 4.3e-05 & 4.0e-05 & 3.8e-05 & 2.7e+03 & 3.2e-05 & 4.3e-05 & 4.0e-05 & 2.4e+03\\

 & \multirow[t]{-2}{*}{\raggedleft\arraybackslash 0.10} & 5.25 & 1.3e-06 & 5.0e-07 & 5.0e-07 & 1.9e+05 & 5.8e-07 & 8.9e-07 & 9.6e-07 & 8.2e+04 & 4.1e-07 & 7.4e-07 & 6.9e-07 & 7.0e+04\\

 &  & 4.35 & 8.8e-05 & 5.0e-04 & 4.9e-04 & 2.2e+02 & 4.3e-05 & 3.8e-04 & 3.8e-04 & 2.3e+02 & 3.2e-05 & 2.3e-04 & 2.4e-04 & 4.6e+01\\

\multirow[t]{-6}{*}{\raggedright\arraybackslash RPS26} & \multirow[t]{-2}{*}{\raggedleft\arraybackslash 0.01} & 5.25 & 1.3e-06 & 6.1e-05 & 6.3e-05 & 1.4e+03 & 5.8e-07 & 3.8e-05 & 3.8e-05 & 1.7e+03 & 4.1e-07 & 2.1e-05 & 2.0e-05 & 2.6e+02\\
\cmidrule{1-15}
 &  & 4.35 & 9.6e-05 & 1.6e-05 & 1.8e-05 & 1.1e+04 & 1.1e-04 & 2.0e-04 & 1.9e-04 & 7.5e+01 & 5.1e-05 & 5.2e-05 & 4.9e-05 & 2.7e+03\\

 & \multirow[t]{-2}{*}{\raggedleft\arraybackslash 0.50} & 5.25 & 1.4e-06 & 9.6e-08 & 8.1e-08 & 1.3e+06 & 1.5e-06 & 5.0e-06 & 4.7e-06 & 1.7e+03 & 7.1e-07 & 7.0e-07 & 6.8e-07 & 1.4e+05\\

 &  & 4.35 & 9.6e-05 & 4.5e-05 & 4.5e-05 & 3.6e+03 & 1.1e-04 & 2.9e-04 & 3.0e-04 & 7.3e+01 & 5.1e-05 & 6.0e-05 & 6.5e-05 & 2.1e+03\\

 & \multirow[t]{-2}{*}{\raggedleft\arraybackslash 0.10} & 5.25 & 1.4e-06 & 8.3e-07 & 8.2e-07 & 1.4e+05 & 1.5e-06 & 1.1e-05 & 1.1e-05 & 8.9e+02 & 7.1e-07 & 9.7e-07 & 9.7e-07 & 8.8e+04\\

 &  & 4.35 & 9.6e-05 & 1.0e-03 & 1.0e-03 & 1.5e+02 & 1.1e-04 & 1.8e-03 & 1.9e-03 & 2.2e+01 & 5.1e-05 & 1.9e-04 & 1.9e-04 & 4.8e+02\\

\multirow[t]{-6}{*}{\raggedright\arraybackslash TPT1} & \multirow[t]{-2}{*}{\raggedleft\arraybackslash 0.01} & 5.25 & 1.4e-06 & 1.4e-04 & 1.3e-04 & 9.4e+02 & 1.5e-06 & 2.5e-04 & 2.5e-04 & 9.5e+01 & 7.1e-07 & 8.8e-06 & 8.6e-06 & 6.0e+03\\
\cmidrule{1-15}
 &  & 4.35 & 4.7e-04 & 2.5e-04 & 2.5e-04 & 3.7e+02 & 4.6e-04 & 3.2e-04 & 3.2e-04 & 3.7e+02 & 4.3e-04 & 4.0e-04 & 3.9e-04 & 3.5e+02\\

 & \multirow[t]{-2}{*}{\raggedleft\arraybackslash 0.50} & 5.25 & 7.1e-06 & 2.5e-06 & 2.4e-06 & 2.2e+04 & 6.9e-06 & 4.0e-06 & 4.0e-06 & 2.0e+04 & 6.4e-06 & 5.7e-06 & 5.7e-06 & 2.0e+04\\

 &  & 4.35 & 4.7e-04 & 5.9e-04 & 6.2e-04 & 1.2e+02 & 4.6e-04 & 5.0e-04 & 5.3e-04 & 1.7e+02 & 4.3e-04 & 4.3e-04 & 4.3e-04 & 2.9e+02\\

 & \multirow[t]{-2}{*}{\raggedleft\arraybackslash 0.10} & 5.25 & 7.1e-06 & 1.8e-05 & 1.8e-05 & 1.4e+03 & 6.9e-06 & 1.2e-05 & 1.2e-05 & 3.3e+03 & 6.4e-06 & 7.3e-06 & 7.2e-06 & 1.4e+04\\

 &  & 4.35 & 4.7e-04 & 6.7e-03 & 6.6e-03 & 1.4e+01 & 4.6e-04 & 3.4e-03 & 3.5e-03 & 2.6e+01 & 4.3e-04 & 9.4e-04 & 9.4e-04 & 1.2e+02\\

\multirow[t]{-6}{*}{\raggedright\arraybackslash CDKN1A} & \multirow[t]{-2}{*}{\raggedleft\arraybackslash 0.01} & 5.25 & 7.1e-06 & 1.1e-03 & 1.0e-03 & 6.4e+01 & 6.9e-06 & 3.9e-04 & 3.8e-04 & 1.5e+02 & 6.4e-06 & 3.9e-05 & 3.9e-05 & 1.8e+03\\
\bottomrule
\end{tabular}
}
\end{table}

\begin{figure}
\centering
\includegraphics[width=0.90\linewidth]{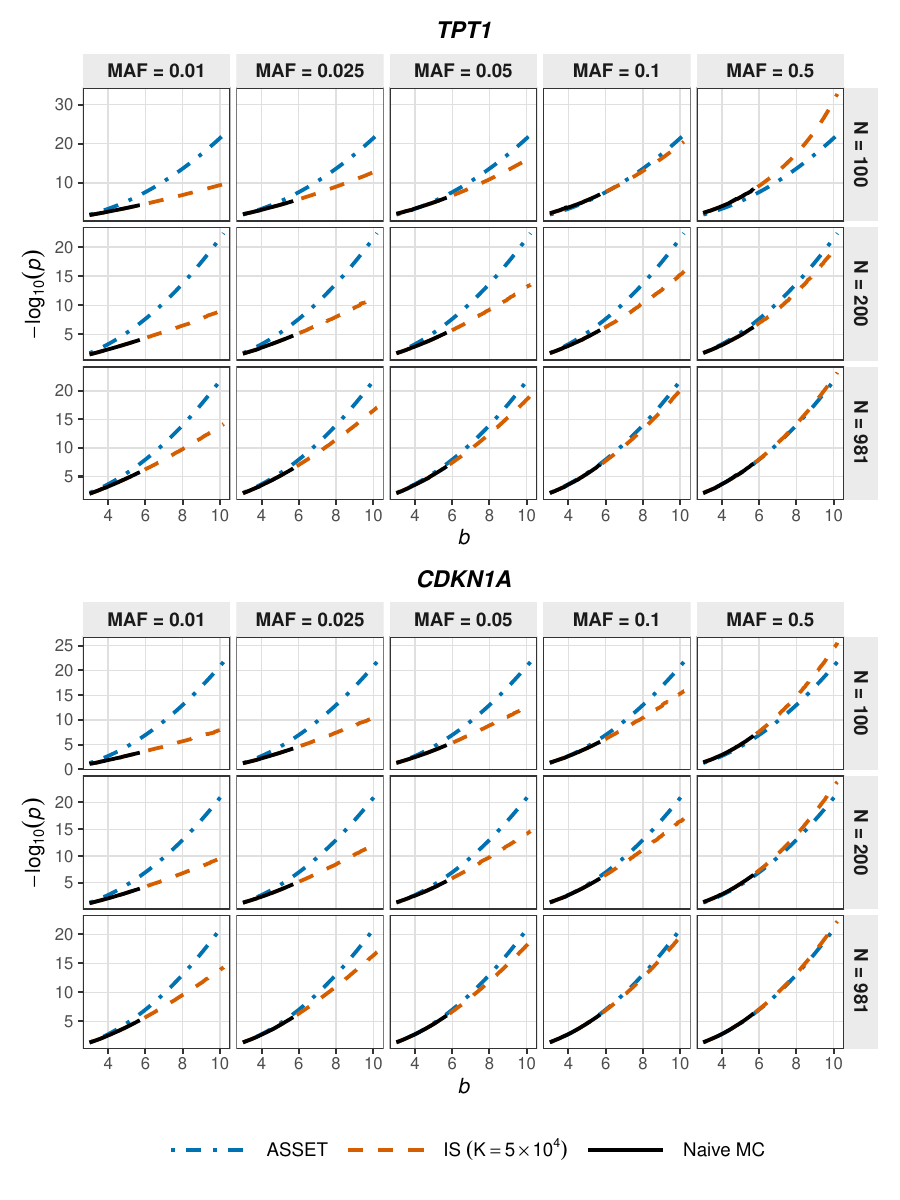}
\caption{Curves of $-\log_{10}\{p(b \mid Y)\}$ versus $b$ for (a) \texttt{TPT1} and (b) \texttt{CDKN1A} from the oneK1K data. Rows correspond to subsample sizes ($N=100, 200$) and the full dataset ($N=981$), and columns correspond to allele frequencies $\mathrm{MAF}\in\{0.01, 0.05, 0.10, 0.25, 0.50\}$. Estimates from importance sampling (IS), ASSET, and naïve Monte Carlo (MC) are shown. IS uses a single tilting parameter within each neighborhood of $b$ values.
}
\label{fig: oneK1K_TPT1_CDKN1A}
\end{figure}

\section{Application to a single-cell eQTL study of lung tissues}\label{sec: real_data}
We applied the IS algorithm to a single-cell eQTL dataset comprising 129 Korean women never-smokers with epithelial cell enrichment, to illustrate the benefit of the IS approach in an independent dataset with a smaller sample size. In the primary analysis, \cite{luong2026single} used TensorQTL \citep{taylor2019scaling} genome-wide significant p-values in each cell type to identify 2,229 eGenes with variants associated with expression in at least one cell type. Here, we applied ASSET to test the global null hypothesis for the corresponding 2,229 gene--SNP pairs identified in the original analysis; the analysis was restricted to SNPs with MAF $> 0.05$ to ensure adequate power. The goal of this application was not to discover additional eGenes, but to compare p-values obtained from the ASSET analytic approximation with those estimated using IS.

In the original analysis, the authors fit linear regression models to assess the association between log-CPM-transformed gene expression and each cis-variant, adjusting for age, the top three genotype principal components, and 25 PEER factors. For each gene--SNP pair, we derived cell-type-specific $z$-scores, $z_c = \hat{\beta}_c / \widehat{\mathrm{se}}(\hat{\beta}_c)$, across the $C = 7$ most abundant cell types available in this dataset. The standardized expression matrix $Y$ was used to estimate the empirical conditional correlation matrix across cell types and to compute the ASSET statistic for each gene--SNP pair. For each observed ASSET statistic, p-values were computed using both the analytic approximation implemented in ASSET and the IS algorithm ($5 \times 10^4$ simulations), conditional on $Y$ and the variant's MAF.

\begin{figure}
\centering
\includegraphics[width=1.0\linewidth]{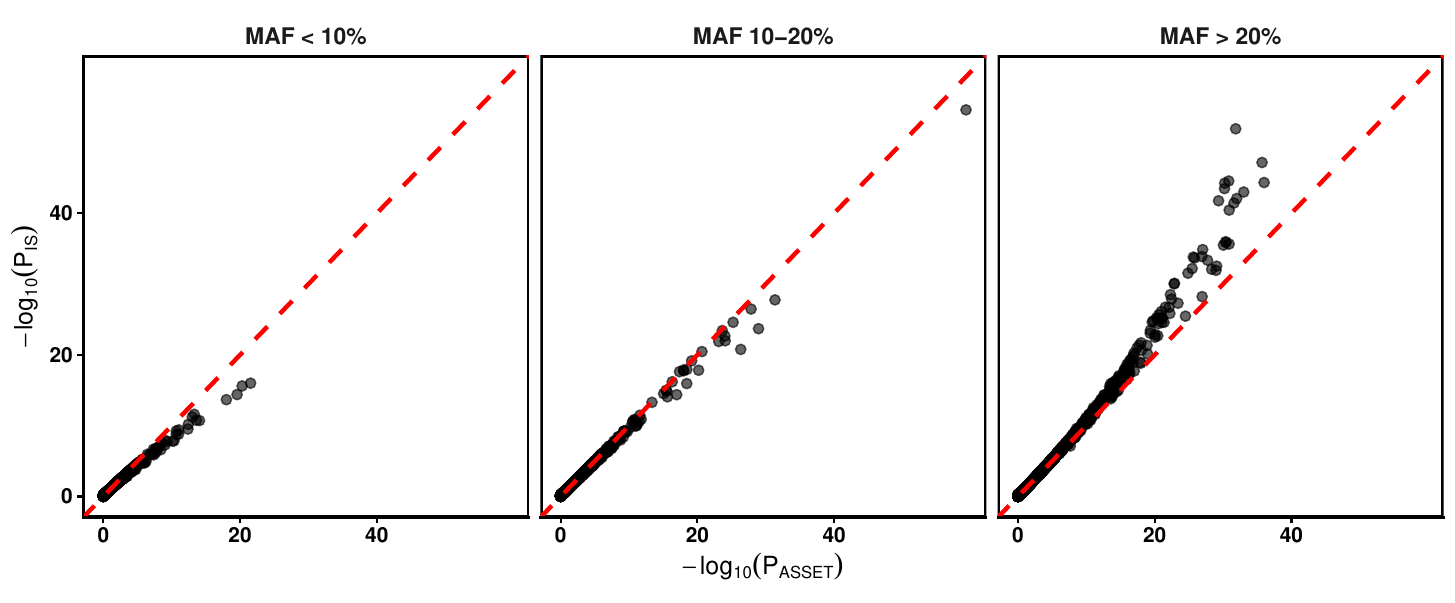}%
\caption{Scatter plot of $-\log_{10}\{p(b)\}$ comparing the conditional IS p-values and the theoretical ASSET (DLM) p-values for the observed ASSET statistic $Z_{\mathrm{ASSET}}=\max_A |Z_A|$. Points are faceted by minor allele frequency (MAF) categories: $\mathrm{MAF}<10\%$, $10\%\le\mathrm{MAF}\le20\%$, and $\mathrm{MAF}>20\%$. The red dashed line denotes the 45$^\circ$ line of equality.}
\label{fig: JCsc-eQTL}
\end{figure}

The comparison of $-\log_{10}$ p-values from ASSET and IS for these gene--SNP pairs is shown in Figure~\ref{fig: JCsc-eQTL}, stratified by MAF ($0.05$--$0.1$, $0.1$--$0.2$, and $> 0.2)$. For variants with $0.05 < \mathrm{MAF} \le 0.1$, signals were generally modest due to limited power, and the two methods agreed closely for most pairs. However, the ASSET analytic approximation tended to produce anti-conservative p-values relative to IS, with discrepancies ranging from negligible (less than $10^{-4}$ on the $\log_{10}$ scale) to approximately 5 orders of magnitude in a small number of cases. Since variants with $\mathrm{MAF} < 0.05$ were excluded from this analysis, the discrepancies were less pronounced than those observed in simulations, which included MAFs as low as 0.01. In contrast, for common variants ($\mathrm{MAF} > 0.2$), the ASSET analytic approximation appeared conservative, consistent with the simulation results.

\section{Discussion}\label{sec: discussion}

ASSET has been widely used to detect modest signals by adaptively aggregating evidence and to explore heterogeneity across multiple related studies. The overall significance of the ASSET statistic is approximated analytically by $p_{\mathrm{DLM}}$, derived using the DLM method under the assumption that test statistics $(z_1,\ldots, z_M)$ are normally distributed. However, the accuracy of this approximation has not been systematically evaluated, particularly for very small p-values or in settings where the normality assumption is questionable. In this manuscript, we present an IS algorithm for estimating the p-value of the ASSET statistic, applicable both when test statistics are normally distributed and when normality may be violated, such as in studies with small sample sizes (e.g., single-cell eQTL studies, rare diseases or subtypes), low-frequency variants, or non-normally distributed phenotypes. The IS algorithm is also applicable to both independent studies and studies with overlapping samples.

Across all scenarios, extensive simulations demonstrate that the proposed IS algorithm agrees closely with naïve Monte Carlo simulations, confirming its accuracy. When the test statistics $(z_1,\ldots, z_M)$ are normally distributed, we find that the analytic p-value $p_{\mathrm{DLM}}$ implemented in the ASSET package is highly accurate across the full range of p-values, as confirmed by comparison with the IS results. This is somewhat surprising, given that $p_{\mathrm{DLM}}$ is theoretically a lower bound on the true p-value. We further show, in the context of single-cell eQTL mapping, that the accuracy of the p-values obtained from ASSET is strongly influenced by allele frequency, phenotype distribution, and sample size. When sample sizes are relatively small, allele frequency is low, or the phenotype distribution is non-normal, $p_{\mathrm{DLM}}$ can yield either anti-conservative or conservative type I error rates, with discrepancies potentially spanning orders of magnitude. Finally, the IS approach offers substantial computational gains over direct Monte Carlo methods, especially for estimating very small p-values. Notably, our simulations indicate that a few hundred IS samples are sufficient to obtain highly accurate estimates, and even as few as 20 samples can produce the correct order of magnitude for extremely small p-values.

There is a long history of using IS to efficiently estimate probabilities of rare events \citep{kahn1949stochastic, goertzel1951monte, kahn1953methods, siegmund1976importance}; see also the introductions in \citep{bucklew2004introduction, liu2001monte}. Naturally, IS has also been applied to p-value estimation in the context of hypothesis testing \citep{siegmund1976importance, naiman2001computing, wu2005p, lloyd2012computing}. In genetics, applications of IS include p-value estimation in linkage studies \citep{malley2003comprehensive, angquist2004using, shi2007importance} and GWAS \citep{kimmel2006fast}. Recent methodological advances in IS can be found in \cite{elvira2021advances}, particularly multiple IS and adaptive IS. A more general IS approach coupled with MCMC \citep{yu2011efficient} was proposed to approximate p-values by partitioning the domain of the test statistic to facilitate efficient sampling; however, this approach requires monitoring convergence. A key step in designing an effective IS algorithm is constructing a proposal distribution that reduces the variance of the IS estimator, either through independent or sequential sampling \citep{liu1998sequential}. In this work, we adopt a proposal measure $P_{\theta}$ that generates independent samples by tilting the sampling distribution toward the rare event $\{\max_A |Z_A| > b\}$ for large $b$, following the seminal work of \cite{siegmund1976importance} in sequential analysis. In practice, the efficiency of the IS estimator is reasonably robust to the choice of $\theta$ \citep{shi2007importance}, allowing efficient estimation of $\log p$ over a range of $b$ using samples generated from $P_{\theta}$ for a fixed $\theta$ (Figure~1). Furthermore, in the context of single-cell eQTL studies, adjusting for PC scores affects only the calculation of the cumulant generating function $\phi_A(\theta)$ but not the simulation of genotypes. Therefore, omitting the PC adjustment in the IS sampling step does not compromise the accuracy of the estimator, although it may slightly reduce efficiency.

In this paper, we focus on the one-sided test, which assumes that all non-null studies have effects in the same direction. For heterogeneous traits, however, genetic effects may differ in direction across studies. \citet{bhattacharjee2012subset} proposed a two-sided ASSET test by deriving the strongest positive and negative associations, denoted by $Z_{\max,+}$ and $Z_{\max,-}$, along with their corresponding p-value approximations $p_{\mathrm{DLM},+}$ and $p_{\mathrm{DLM},-}$, obtained by conditioning on the signs of the test statistics. When studies are independent, $p_{\mathrm{DLM},+}$ and $p_{\mathrm{DLM},-}$ are independent, and Fisher’s method can be used to combine them to derive a final p-value. Similarly, under independence, a slightly modified IS algorithm can be used to estimate $p_{\mathrm{DLM},+}$ and $p_{\mathrm{DLM},-}$, and then combine them to obtain the overall p-value. However, this strategy does not extend directly to settings with dependent studies, where complex correlations arise between $p_{\mathrm{DLM},+}$ and $p_{\mathrm{DLM},-}$. This is left as future work; one possible approach is to use the aggregated Cauchy association test \citep{liu2019acat}, which does not require explicit specification of the correlation structure between the two p-values.

Finally, we note that combining evidence across multiple studies or traits to detect modest signals is a widely used strategy. Classical methods include Fisher's combined p-value test \citep{fisher1925statistical} and several extensions designed to maintain power under heterogeneity, such as the truncated product method \citep{zaykin2002truncated}, the adaptive rank truncated product (ARTP) \citep{yu2009pathway}, and the adaptively weighted Fisher statistic \citep{li2011adaptively}. While Fisher's method has a known analytic null distribution, these adaptive extensions typically require extensive simulation to evaluate significance and do not directly assess effect heterogeneity. In contrast, ASSET provides a highly accurate analytic approximation under normality and offers direct interpretability of which subset of studies drives the association. On the Bayesian side, methods such as mashr \citep{urbut2019flexible} aggregate modest associations across studies by learning a prior on effect sizes and borrowing information across traits to improve effect-size estimation. Such approaches have proved particularly useful in single-cell and multi-tissue eQTL analyses for characterising effect-size patterns across cell types of varying statistical power, enabling principled comparisons that account for differences in sample composition and measurement variability \citep{yazar2022single, natri2024cell}.

\subsection*{Acknowledgment}
This research was supported by the Intramural Research Program of the National Institutes of Health (NIH). The contributions of the NIH author(s) are considered Works of the United States Government. The findings and conclusions presented in this paper are those of the author(s) and do not necessarily reflect the views of the NIH or the U.S. Department of Health and Human Services. This study utilized the high-performance computational capabilities of the Biowulf Linux cluster at the National Institutes of Health, Bethesda, MD. (\href{https://hpc.nih.gov/}{https://hpc.nih.gov/}). 

 
\subsection*{Competing interests}
The authors declare no conflict of interest. 

\subsection*{Code availability}
The developed software and the test data are available from GitHub at \href{https://github.com/samuelanyaso/assetIS}{https://github.com/samuelanyaso/assetIS}


\end{document}